\documentclass[reprint,amsmath,amssymb,pra,aps,twocolumn,showpacs]{revtex4-1}

\usepackage{mathrsfs}
\usepackage{graphicx}
\usepackage{amsmath}
\usepackage{amssymb}
\usepackage{amsfonts}
\usepackage{color}
\usepackage{CJK}
\usepackage{leftidx}

\def\bra#1{\left\langle{#1}\right|}
\def\ket#1{\left|{#1}\right\rangle}
\def\braket#1#2{\left\langle{{#1}}\mathrel{\left|{\vphantom{{#1}{#2}}}\right.\kern-\nulldelimiterspace}{{#2}}\right\rangle}

\begin{document}

\title{Multiparameter estimation via an ensemble of spinor atoms}

\author{Min Zhuang$^{1,2}$}

\author{Jiahao Huang$^{1}$}
\altaffiliation{Email: hjiahao@mail2.sysu.edu.cn, eqjiahao@gmail.com}

\author{Chaohong Lee$^{1,2}$}
\altaffiliation{Email: lichaoh2@mail.sysu.edu.cn, chleecn@gmail.com}

\affiliation{$^{1}$Laboratory of Quantum Engineering and Quantum Metrology, School of Physics and Astronomy, Sun Yat-Sen University (Zhuhai Campus), Zhuhai 519082, China}

\affiliation{$^{2}$State Key Laboratory of Optoelectronic Materials and Technologies, Sun Yat-Sen University (Guangzhou Campus), Guangzhou 510275, China}

\begin{abstract}
Multiparameter estimation, which aims to simultaneously determine multiple parameters in the same measurement procedure, attracts extensive interests in measurement science and technologies.
Here, we propose a multimode many-body quantum interferometry for simultaneously estimating linear and quadratic Zeeman coefficients via an ensemble of spinor atoms.
Different from the scheme with individual atoms, by using an $N$-atom multimode GHZ state, the measurement precisions of the two parameters can  simultaneously attain the Heisenberg limit, and they respectively depend on the hyperfine spin number $F$ in the form of  $\Delta p \propto 1/(FN)$ and $\Delta q \propto 1/(F^2N)$.
Moreover, the simultaneous estimation can provide better precision than the individual estimation.
Further, by taking a three-mode interferometry with Bose condensed spin-1 atoms for an example, we show how to perform the simultaneous estimation of $p$ and $q$.
Our scheme provides a novel paradigm for implementing multiparameter estimation with multimode quantum correlated states.
\end{abstract}

\date{\today}

\maketitle

\section{Introduction}\label{Sec1}
Precision metrology and parameter estimation are of great importance in both fundamental sciences and practical technologies~\cite{VGiovannetti2004,VGiovannetti2006,VGiovannetti2011,NHinkley2013,Martin2013}.
Quantum metrology has opened up a new frontier of high-precision parameter estimation by using quantum strategies rather than classical ones~\cite{IDLeroux2010,CGross2010,Blucke2011,JHuang2014,WMuessel2015}.
The two-mode quantum interferometry had been widely used for single-parameter estimation~\cite{SLBraunstein1994,GTOth2014,RDemkowicz2015}.
According to the central limit theorem, the parameter measurement precision via $N$ individual particles is scaled as the standard quantum limit (SQL), i.e., $\propto1/\sqrt{N}$.
Nevertheless, by utilizing entangled particles, the SQL can be surpassed.
In particular, by inputing Greenberger-Horne-Zeilinger (GHZ) state or NOON state, the parameter measurement precision can be scaled as the Heisenberg limit, i.e., $\propto1/N$~\cite{JJBollinger1996,CLee2006,TMonz2011,CLee2012,JHuang2015,SDHuver2008}.

Despite single-parameter estimation plays an important role in many aspects, the realistic problems may generally involve multiple parameters.
Therefore, estimating multiple parameters simultaneously becomes an important task in quantum metrology~\cite{MSzczykulska2016,TJProctor2018,MGessner2018}, such as microscopy, spectroscopy~\cite{TMonz2011,LeibfriedD2004}, electromagnetic field sensing~\cite{JMTaylor2008,JRMaze2008,BJMaertz2010,GdeLangez2011}, gravitational field detection~\cite{CMCaves1981,AAcin2001,RXAdhikari2014} and so on.
Recently, the studies on simultaneous quantum-enhanced estimation of multiple parameters have attracted great interests.
Under equivalent quantum resources, simultaneous estimation of multiple parameters may yield better precision than estimating them individually.
It had been demonstrated that, multiparticle entanglement can provide an enhancement for the multiple phase imaging~\cite{NSpagnolo2012,PCHumphreys2013,JD2016,MACiampini2016,PAKnott2016,CNGagatsos2016,JLiu2016,JRehacek2017,LZhang2017} and multidimensional field estimation~\cite{TBaumgratz2016}.
Besides, two-mode entangled coherent states are proposed for simultaneously estimating the linear and nonlinear phase shifts~\cite{JCheng2014}.

Bose condensed spinor atoms, which involve multiple spin degrees of freedom~\cite{YKawaguchia2012}, provide an excellent candidate for demonstrating multiparameter estimation.
In the presence of magnetic field, the Zeeman states $|F,m_F\rangle$ are split separately and then can act as multiple paths or multiple modes.
Usually, when the magnetic field strength is weak, the linear Zeeman effect dominates.
When the magnetic field strength is sufficiently strong, the quadratic Zeeman effect governs.
For intermediate magnetic field strength, the Zeeman splitting is compared to the hyperfine energy splitting, the linear Zeeman term $E_\textrm{{LZ}}\propto{p m_F}$ and quadratic Zeeman term $E_\textrm{{NLZ}}\propto{q m_F^2}$ are in competition. Here, $m_F$ is the magnetic quantum number, $p$ and $q$ are coefficients for linear and quadratic Zeeman effects, respectively.
Thus, in this scenario, simultaneously estimating the linear and quadratic Zeeman coefficients $p$ and $q$ becomes an interesting problem.
Naturally, the following questions arise: i) can one use multimode many-body quantum states to simultaneously estimate these parameters? ii) what are their measurement precision bounds? and iii) is it possible to realize in a state-of-the-art experiment?

In this paper, by considering an ensemble of Bose condensed spinor atoms, we propose a multimode many-body quantum interferometry for simultaneously estimating the linear and quadratic Zeeman coefficients.
We find that, if the atoms are prepared in a multimode GHZ state, the measurement precisions of the two parameters can attain the Heisenberg limit simultaneously, and they respectively depend on the hyperfine spin number $F$ according to $\Delta p \propto 1/(FN)$ and $\Delta q \propto 1/(F^2N)$.
The simultaneous estimation provides better precision than the individual strategy.
Further, we take a three-mode interferometry with spin-1 Bose condensed atoms for example and show how to perform the multiparameter estimation in experiments.
Our scheme is a simple multiparameter estimation scenario with multimode many-body quantum states. It may provide a novel paradigm for implementing quantum sensing via multimode quantum correlated states.

The paper is organised as follows.
In Sec.~\ref{Sec2}, we introduce the scheme of multimode interferometer via spin-F Bose condensed atoms for simultaneous estimation of $p$ and $q$.
In Sec.~\ref{Sec3}, we discuss the ultimate bounds for the two parameters via individual atoms as well as entangled atoms.
%
%
More importantly, input a specific $N$-atom multimode GHZ state, the measurement precision of $p$ and $q$ can reach the Heisenberg limit simultaneously.
In Sec.~\ref{Sec4}, we proceed with the comparisons of the schemes on simultaneous and individual estimation. The measurement precisions via simultaneous estimation are found to be advantageous.
In Sec.~\ref{Sec5}, as an example, we propose the practical three-path interferometric schemes via spin-1 Bose condensed atoms. We show how to estimate $p$ and $q$ via observable measurements.
Finally, we give a brief summary in Sec.~\ref{Sec6}.

\section{Two-parameter estimation via quantum interferometry of spinor atoms}\label{Sec2}

\begin{figure}[!htp]
 \includegraphics[width=1\columnwidth]{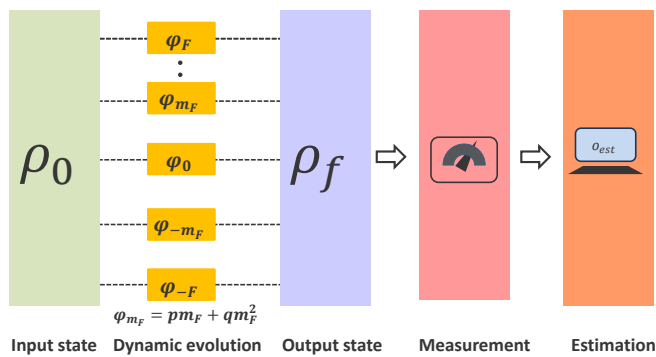}
  \caption{\label{Fig1}(color online) Schematic of the multiparameter estimation with spin-$F$ atoms. The spin-$F$ atoms contains $2F+1$ Zeeman sublevels ($\ket{F,m_F}$, $m_F=-F,-F+1,...,F$), and a multiparticle state involved $2F+1$ Zeeman sublevels can be prepared as the input state. Then, the input state will accumulate different phases dependent on the two unknown parameters $p$ and $q$ caused by linear and quadratic Zeeman effects respectively. Finally, the measurement onto the output state and the parameter estimation are implemented.}
\end{figure}

Based upon an ensemble of spin-$F$ atoms in an external magnetic field, we introduce our interferometric scheme for simultaneously estimating the linear Zeeman coefficient $p$ and the quadric Zeeman coefficient $q$.
Different from the conventional two-mode interferometry~\cite{UDorner2009,RDemkowicz2009,BMEscher2011,CLuo2017}, our scheme extends to a multi-mode interferometer, where the Zeeman sublevels of the spinor atoms act as the interferometry modes.
Thus, in an ensemble of spin-$F$ atoms, there are $2F+1$ Zeeman sublevels denoted by $\ket{F,m_F}$, with $m_F=-F,-F+1,...,F$ the magnetic quantum number.
The Hamiltonian of $N$ spin-$F$ atoms in a magnetic field reads (we set $\hbar=1$ throughout the paper),
\begin{equation}\label{Eq:Hamiltonian}
 \hat{H}{(p,q)}=\sum\limits_{n=1}^Np\hat{s}^{[n]}_z+q\hat{s}^{2 [n]}_z,
\end{equation}
where $p=|g|\mu_B B$ and $q=\frac{(g\mu_\mathrm{B}B)^2}{\Delta\mathrm{E}_\mathrm{hf}}$ denote the linear Zeeman coefficient and the quadric Zeeman coefficient respectively.
$B$ is the external magnetic field that is assumed to be applied in the $z$ direction, $\mu_B$ is the Bohr magneton, $\Delta\mathrm{E}_\mathrm{hf}$ is the hyperfine energy spitting, g is the land$\acute{e}$ hyperfine g-factor.
$\hat{s}_z^{[n]}$ is spin-$F$ magnetic angular momentum operator for the $n$-th atom with $n=1,2,\ldots,N$.

The procedures of our scheme are as follows, see Fig.~\ref{Fig1}.
First, an $N$-particle state $\ket{\Psi}$ involved $2F+1$ Zeeman sublevels is prepared as the input probe state.
Then, the multi-path input probe state acquires phase shifts depending on the corresponding Zeeman sublevels,  $\ket{\Psi(p,q)}=\hat{U}(p,q)\ket{\Psi}$,
and thus the information of parameters $p$ and $q$ are imprinted in the evolved output state $ \ket{\Psi(p,q)}$.
The phase-imprinted dynamical evolution can be described by
\begin{equation}\label{1}
  \hat{U}(p,q)=\mathit{e}^{-i\hat{H}(p,q)T}
\end{equation}
where $T$ is the evolution duration and different Zeeman sublevels accumulate different phase $\varphi_{m_F}(p,q)$ during the same evolution.
The Zeeman sublevel $\ket{F,0}$ acts the role of reference mode [$\varphi_{0}(p,q)=0$], and the other Zeeman sublevels $\ket{F,m_F}$ register the relative phase $\varphi_{m_F}(p,q)$.
Finally, the unknown parameters $p$ and $q$ is extracted by inferring the information of the output state.

\section{Measurement precision bounds}\label{Sec3}

In this section, we discuss the ultimate measurement precision bounds for the two parameters via our multimode interferometry with different input states.
Without loss of generality, we assume the evolution duration $T=1$ in our calculation.
In Sec.~\ref{Sec3a} and Sec.~\ref{Sec3b}, we show the ultimate bounds for the two parameters via individual and entangled Bose condensed spin-$F$ atoms, respectively.
%
%
%

According to the multiparameter quantum estimation theory~\cite{CWHelstrom1967,CWHelstrom1976,MGAParis2009},
the precision of the two parameters is determined by its covariance matrix $\mathrm{Cov}(p,q)$,
and the lower bound of the $\mathrm{Cov}(p,q)$ [the quantum Cram\'{e}r-Rao bound (QCRB)] is characterized by the quantum Fisher information matrix (QFIM),
\begin{equation}\label{Eq:Cov-FQ}
  \mathrm{Cov}(p,q)\geq(\mu[\mathnormal{\mathbf{F}}_Q(p,q)]^{-1}),
\end{equation}
where $\mu$ is the number of the experiment trials, $[\mathnormal{\mathbf{F}}_Q(p,q)]^{-1}$ is inverse matrix of the QFIM.
The matrix element of QFIM is written as,
\begin{equation}\label{Eq:FQ}
 [\mathnormal{\mathbf{F}}_Q(p,q)]_{k,l}=\mathrm{Tr}\left[\rho(p,q)\left(\hat{L}_k\hat{L}_l+\hat{L}_k\hat{L}_l\right)/{2}\right],
\end{equation}
\begin{equation}
 (k,l=1,2) \nonumber
\end{equation}
where $\hat{L}_k$ is symmetric logarithmic derivatives (SLD) and $\rho(p,q)$ is the density matrix of a system state.
For a pure input state and unitary evolution, the $\hat{L}_k$ can be explicitly expressed as
\begin{eqnarray}\label{Eq:SLD_q}
  \hat{L}_1=2(\ket{\partial_{p}\Psi(p,q)}\bra{\Psi(p,q)}+\ket{\Psi(p,q)}\bra{\partial_{p}\Psi(p,q)}),\nonumber\\
  \hat{L}_2=2(\ket{\partial_{q}\Psi(p,q)}\bra{\Psi(p,q)}+\ket{\Psi(p,q)}\bra{\partial_{q}\Psi(p,q)}),\nonumber\\
\end{eqnarray}
where $\ket{\partial_{p}\Psi(p,q)}\!=\!\!\partial_{p}\hat{U}(p,q)\ket{\Psi}$ and $\ket{\partial_{q}\Psi(p,q)}\!=\!\!\partial_{q}\hat{U}(p,q)\ket{\Psi}$ denote the partial derivatives of $\ket{\Psi(p,q)}$ with respect to the parameter $p$ and $q$.
Thus, the matrix element of QFIM can be simplified as~\cite{TBaumgratz2016}
\begin{equation}\label{Eq:FQN}
  [\mathnormal{\mathbf{F}}_Q(p,q)]_{k,l}=4N\mathbf{I}_{k,l}^{[1]}+4N(N-1)\mathbf{I}_{k,l}^{[2]},
\end{equation}
with
\begin{equation}\label{Eq:Ikl-1}
 \mathbf{I}_{k,l}^{[1]}=\mathrm{Tr}[\hat\rho^{[1]}\hat{a}_k\hat{a}_l]-\mathrm{Tr}[\hat\rho^{[1]}\hat{a}_k]\mathrm{Tr}[\hat\rho^{[1]}\hat{a}_l],
 \nonumber
\end{equation}
\begin{equation}\label{Eq:Ikl-2}
 \mathbf{I}_{k,l}^{[2]}=\mathrm{Tr}[\hat\rho^{[2]}\hat{a}_k\otimes\hat{a}_l]-\mathrm{Tr}[\hat\rho^{[1]}\hat{a}_k]\mathrm{Tr}[\hat\rho^{[1]}\hat{a}_l].
 \nonumber
\end{equation}
Here, $\rho^{[1]}$ is the one-particle reduced density matrix of the input state, $\rho^{[2]}$ is the two-particle reduced density matrix of the input state, with $\hat{a}_1=\hat{s}_z,\hat{a}_2=\hat{s}^2_z$.
Since the variance of the two parameters $p$ and $q$ are the diagonal terms of the covariance matrix $\mathrm{Cov}(p,q)$,
then the uncertainties of the two parameters are limited by the QCRB,
\begin{eqnarray}\label{Eq:Delta_p_q_general}
 \Delta^2{p}\geq{{\Delta^2{p}}}_\textrm{{QCRB}}\equiv[\mathnormal{\mathbf{F}}_Q(p,q)]^{-1}_{11},
 \nonumber\\
 \Delta^2{q}\geq{\Delta^2{q}}_\textrm{{QCRB}}\equiv[\mathnormal{\mathbf{F}}_Q(p,q)]^{-1}_{22}.
\end{eqnarray}
\subsection{Measurement precision bounds offered by individual atoms \label{Sec3a}}

%
We first consider individual atoms without any entanglement.
%
%
Thus, the system state can be described by a product state of $N$ atoms, which is written as
\begin{equation}\label{Eq:productstate}
 \ket{\Psi}^\textrm{{Pro}}=\left(\sum\limits_{m_F=-F}^{F}\!\!\alpha_{m_F}\ket{F,m_F}\right)^{\otimes N},
\end{equation}
where $\alpha_m$ is the complex amplitude and satisfies the normalization condition
$\sum\limits_{m_F=-F}^{F}\!\!|\alpha_{m_F}|^2=1$.
For input product states, due to $\rho^{[2]}=\rho^{[1]}\otimes\rho^{[1]}$, the second term of Eq.~\eqref{Eq:FQN} is zeros, and thus the QFIM only scales linearly with $N$.
Take the $N$-atom product state~\eqref{Eq:productstate} as the input state, its QFIM can be calculated,
\begin{eqnarray}\label{Eq:FQM_Product}
\mathbf{F}_Q^{\textrm{{Pro}}}(p,q)=N\mathbf{F}_Q^{S}(p,q),
\end{eqnarray}
where $\mathbf{F}_Q^{S}(p,q)$ denotes the QFIM of one individual atom and its explicit expression can be written as
\begin{eqnarray}\label{Eq:FQM_single}
\mathbf{F}_Q^{S}(p,q)\!\!=\left(
  \begin{array}{cc}
   4(M_2-M_1^2) &4(M_3-M_1M_2) \\
   4(M_3-M_1M_2)&4(M_4-M_2^2) \\
 \end{array}
\right),\nonumber\\
\end{eqnarray}
with
\begin{eqnarray}\nonumber
 M_1\!=\!\sum_{m_F=-F}^{F}\!|\alpha_{m_F}|^2 m_F, \nonumber\\
 M_2\!=\!\sum_{m_F=-F}^{F}\!|\alpha_{m_F}|^2 m_F^2,\nonumber\\
 M_3\!=\!\sum_{m_F=-F}^{F}\!|\alpha_{m_F}|^2 m_F^3, \nonumber\\
 M_4\!=\!\sum_{m_F=-F}^{F}\!|\alpha_{m_F}|^2 m_F^4.
\end{eqnarray}
Therefore, the ultimate bounds for estimating the two parameters with individual atoms are given by
\begin{widetext}
\begin{eqnarray}\label{Eq:deltap_q_Product}
  {\Delta^2{p}}_\textrm{{QCRB}}=[\mathbf{F}_Q^{\textrm{{Pro}}}(p,q)]^{-1}_{11}=\frac{[\mathbf{F}_Q^{S}(p,q)]_{22}}{N([\mathbf{F}_Q^{S}(p,q)]_{11}[\mathbf{F}_Q^{S}(p,q)]_{22}-[\mathbf{F}_Q^{S}(p,q)]^{2}_{12})},\nonumber\\
  {\Delta^2{q}}_\textrm{{QCRB}}=[\mathbf{F}_Q^{\textrm{{Pro}}}(p,q)]^{-1}_{22}=\frac{[\mathbf{F}_Q^{S}(p,q)]_{11}}{N([\mathbf{F}_Q^{S}(p,q)]_{11}[\mathbf{F}_Q^{S}(p,q)]_{22}-[\mathbf{F}_Q^{S}(p,q)]^{2}_{12})}.\nonumber\\
\end{eqnarray}
\end{widetext}
It is obvious that, the input state of single atom will affect the value of $\mathbf{F}_Q^{S}(p,q)$ and thus give different measurement precision bound.

If the single atom state is in uniform distribution among the Zeeman sublevels,
\begin{equation}\label{symmetry_state}
  \ket{\psi}_\textrm{{uni}}^{\textrm{Pro}}=\left(\frac{1}{\sqrt{2F+1}}\sum\limits_{m_F=-F}^{F}\!\!\ket{F,m_F}\right)^{\otimes N},
\end{equation}
its QFIM can be obtained analytically, i.e.,
\begin{equation}\label{Eq:F_Q_symmetry}
\mathbf{F}_Q^{S}(p,q)=4N\left(
  \begin{array}{cc}
    \frac{F(1+F)}{3} & 0\\
    0                & \frac{F(1+F)(4F^2+4F-3)}{45}\\
 \end{array}
\right),
\end{equation}
and thus the measurement precision bounds for the two parameters are
\begin{equation}\label{Eq:deltap1symme}
 {\Delta p}_\textrm{{QCRB}}=\sqrt{\frac{3}{4NF(1+F)}},
\end{equation}
and
\begin{equation}\label{Eq:deltap2symme}
 {\Delta q}_\textrm{{QCRB}}=\sqrt{\frac{45}{4NF(1+F)(4F^2+4F-3)}}.
\end{equation}
When $F$ is large, ${\Delta p}_\textrm{{QCRB}}\propto\frac{1}{F}$ while ${\Delta q}_\textrm{{QCRB}}\propto\frac{1}{F^2}$.
\begin{figure}[!htp]
 \includegraphics[width=1\columnwidth]{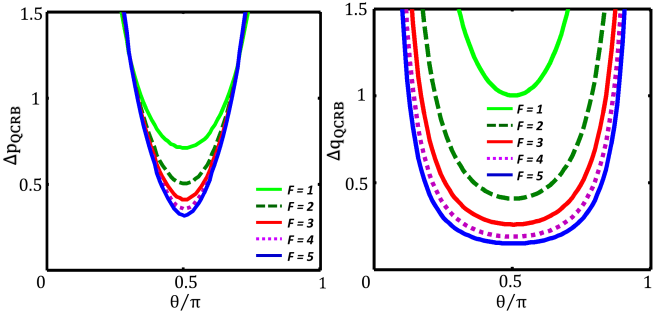}
  \caption{\label{Fig_scs_state_precision_theta}(color online).
  The measurement precision $\Delta {p_\textrm{{QCRB}}}$ and $\Delta {q_\textrm{{QCRB}}}$ versus $\theta$ for different single spin-$F$ atom in binomial distribution among Zeeman sublevels.}
\end{figure}

Then, we consider that the single atom state is in binomial distribution among the Zeeman sublevels, i.e.,
\begin{eqnarray}\label{Eq:spin-coherent-state}
  &&\ket{\psi}_{\textrm{bio}}^{\textrm{Pro}}=((1+\varepsilon\varepsilon^*)^{-F}\ket{\varepsilon})^{\otimes N},\nonumber \\
  &&\ket{\varepsilon}=\!\!\!\sum\limits_{m_F=-F}^{F}\!\sqrt{\frac{2F!}{(F\!-\!m_F)!(F\!+\!m_F)!}}
  \varepsilon^{F\!-\!m_F}\!\ket{F,m_F},
  \nonumber \\
  &&\varepsilon=\tan\frac{\theta}{2}e^{i\varphi},
\end{eqnarray}
where $\varepsilon$ parameterizes the direction through the azimuthal and polar angles $\varphi$ and $\theta$~\cite{JMRadcliffe1971}.
We set $\varphi=0$ in our calculation since $\varphi$ has no effects on the QFIM.
The inverse matrix of QFIM for Eq.~\eqref{Eq:spin-coherent-state} can be calculated analytically, but it is a bit complicated so that we do not show it explicitly here.
Our results demonstrate that, the ${\Delta p}_\textrm{{QCRB}}$ and ${\Delta q}_\textrm{{QCRB}}$ are sensitively dependent the choice of $\theta$.
In Fig.~\ref{Fig_scs_state_precision_theta}, we show the ultimate measurement precision ${\Delta p}_\textrm{{QCRB}}$ and ${\Delta q}_\textrm{{QCRB}}$ versus $\theta$.
For both $p$ and $q$, $\theta=\pi/2$ is the optimal choice where ${\Delta p}_\textrm{{QCRB}}$ and ${\Delta q}_\textrm{{QCRB}}$ attain their minimum simultaneously.

\begin{figure}[!htp]
 \includegraphics[width=1\columnwidth]{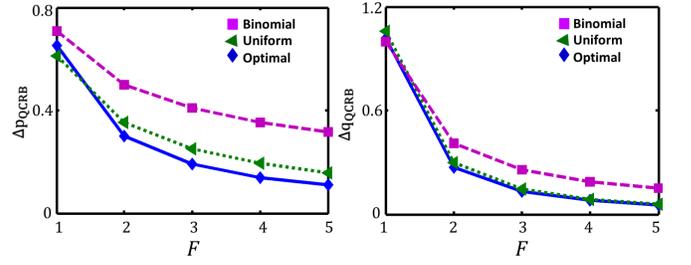}
  \caption{\label{Fig_Single_atom}(color online).
  The measurement precision $\Delta p_{\textrm{{QCRB}}}$ and $\Delta q_{\textrm{{QCRB}}}$ versus with respect to the hyperfine spin $F$ for the three different kinds of input state of the single atom.}
\end{figure}

Further, to find out the optimal distribution among the Zeeman sublevels for estimating the two parameters $p$ and $q$ simultaneously, we minimize the sum of the variance ${\Delta^2{p}}_\textrm{{QCRB}}+\Delta^2{q}_\textrm{{QCRB}}$ by means of numerical methods.
For the $N$-atom product state~\eqref{Eq:productstate} with $F\in{[1,2,3,4,5]}$, the optimal state for individual atoms is always in the form of
%
\begin{equation}\label{Eq:optimal-state-single}
\ket{\psi}_{\textrm{{opt}}}^\textrm{{Pro}}=\left(\alpha_F\ket{F,F}+\alpha_0\ket{F,0}+\alpha_{-F}\ket{F,-F}\right)^{\otimes N},
\end{equation}
where $\alpha_F$, $\alpha_0$, and $\alpha_{-F}$ (satisfying $|\alpha_F|^2=|\alpha_{-F}|^2$ and $|\alpha_{0}|^2>0$) are dependent on the hyperfine spin $F$ and can be determined by numerical calculation.
Given the optimal form of the input state, the corresponding ultimate bounds for the two parameters are
\begin{equation}\label{Eq:deltapoptimal}
 {\Delta p}_\textrm{{QCRB}}=\frac{1}{2\sqrt{2N} |\alpha_F| F},
\end{equation}
and
\begin{equation}\label{Eq:deltap2symme}
 {\Delta q}_\textrm{{QCRB}}=\frac{1}{2\sqrt{2N} |\alpha_F| F^2}.
\end{equation}
The measurement precision of $p$ can have a $F$-fold enhancement while the measurement precision of $q$ can obtain a $F^2$-fold magnification.

For comparison, we show how the measurement precisions ${\Delta p}_\textrm{{QCRB}}$ and ${\Delta q}_\textrm{{QCRB}}$ change with the hyperfine spin $F$ for uniform, binomial ($\theta=\pi/2$) and optimal distribution in Fig.~\ref{Fig_Single_atom}.
It is shown that, the measurement precision of the uniform distribution is close to the optimal one, while the binomial distribution performs a bit worse.
Despite the dependences of the uncertainties on $F$ are quite different for the three distribution, they all decrease monotonously as $F$ increases.
This indicates that, with suitable distribution among Zeeman sublevels, the measurement precision of the two parameters can be improved simultaneously by using the spinor atoms with larger hyperfine spin $F$.
%

\subsection{Measurement precision bounds offered by entangled atoms}\label{Sec3b}

It is well known that, entanglement can improve the measurement precision.
In the following, we show how the entangled input states of $N$ atoms can simultaneously enhance the measurement precisions of the two parameters.
Specifically, we consider a multimode GHZ state as an input state, which is in the form of
\begin{equation}\label{Eq:GHZ}
  \ket{\Psi}^\textrm{{GHZ}}=\sum\limits_{m_F=-F}^{F}\alpha_{m_F}\left(\ket{F,m_F}^{\otimes N}\right).
\end{equation}

Substituting the input state~\eqref{Eq:GHZ} into Eqs.~\eqref{Eq:FQN} and~\eqref{Eq:Delta_p_q_general}, we obtain the corresponding uncertainty bounds for the two parameters
\begin{widetext}
\begin{eqnarray}\label{Eq:deltap_q_GHZ}
 {\Delta^2{p}}_\textrm{{QCRB}}=[\mathbf{F}_Q^\textrm{{GHZ}}(p,q)]^{-1}_{11}=\frac{[\mathbf{F}_Q^{\mathrm{S}}(p,q)]_{22}}{N^2([\mathbf{F}_Q^{\mathrm{S}}(p,q)]_{11}[\mathbf{F}_Q^{\mathrm{S}}(p,q)]_{22}-[\mathbf{F}_Q^{\mathrm{S}}(p,q)]^{2}_{12})}, \nonumber\\
  {\Delta^2{q}}_\textrm{{QCRB}}=[\mathbf{F}_Q^\textrm{{GHZ}}(p,q)]^{-1}_{22}=\frac{[\mathbf{F}_Q^{\mathrm{S}}(p,q)]_{11}}{N^2([\mathbf{F}_Q^{\mathrm{S}}(p,q)]_{11}[\mathbf{F}_Q^{\mathrm{S}}(p,q)]_{22}-[\mathbf{F}_Q^{\mathrm{S}}(p,q)]^{2}_{12})}.\nonumber\\
\end{eqnarray}
\end{widetext}
Comparing the Eq.~\eqref{Eq:deltap_q_GHZ} with Eq.~\eqref{Eq:deltap_q_Product}, we can immediately find that the scaling of the measurement precisions versus total atomic number $N$ change from $1/\sqrt{N}$ to $1/N$.
That is, in principle, the GHZ's form of $N$-atom state can improve the measurement precisions of the two parameters simultaneously to the Heisenberg limit.

\begin{figure}[!htp]
  \includegraphics[width=1\columnwidth]{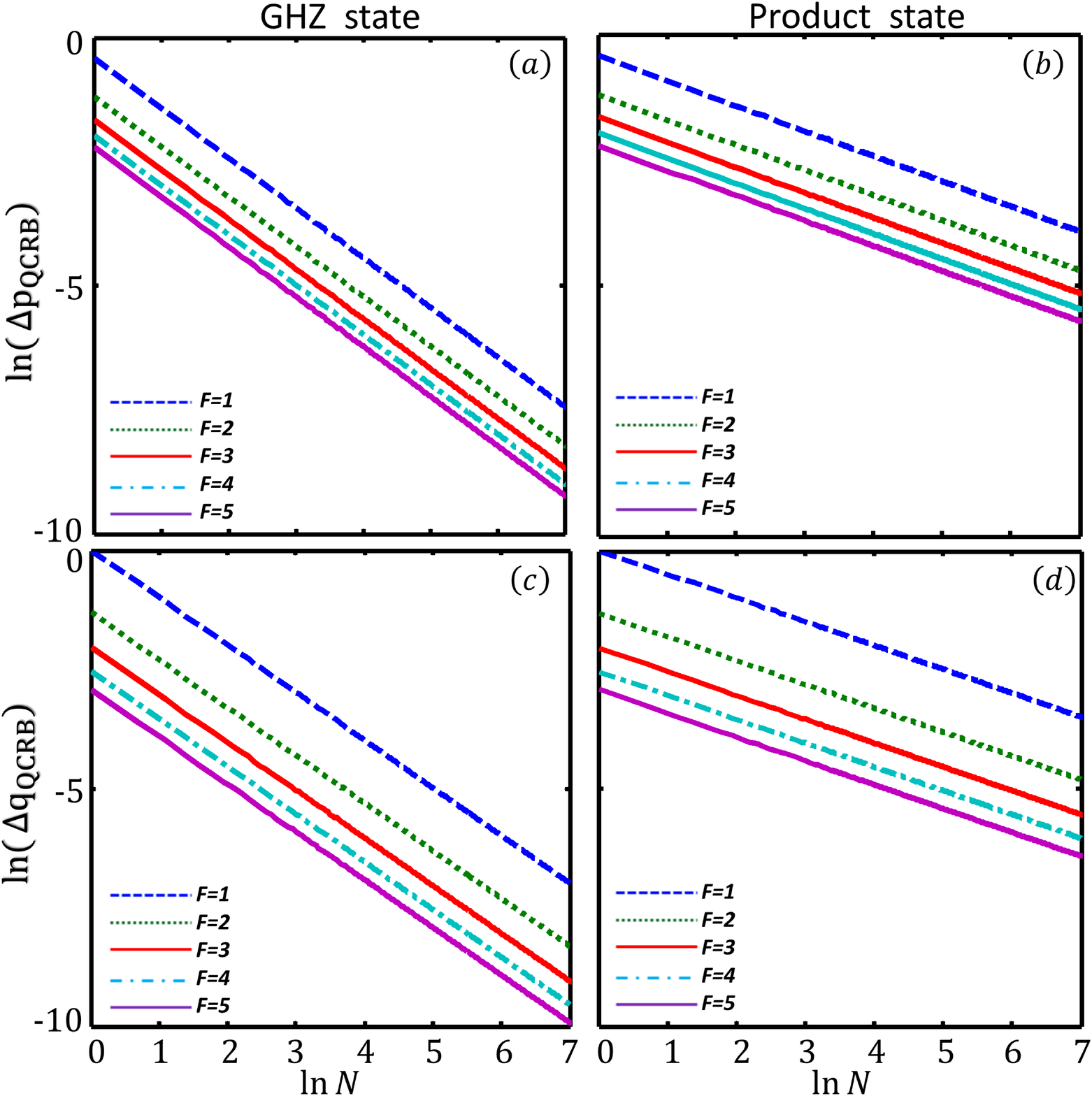}
  \caption{\label{Fig_scaling_optimal_product_and GHZ}(color online).
  The log-log scaling of $\Delta p_\textrm{{QCRB}}$ and ${\Delta q}_\textrm{{QCRB}}$ versus the total atomic number $N$ under different hyperfine spin $F$ for the optimal GHZ state and optimal product state.}
\end{figure}

For the multimode GHZ state~\eqref{Eq:GHZ} with $F\in[1,2,3,4,5]$, the optimal one reads as
\begin{equation}\label{Eq:optimal-state-GHZ}
\ket{\Psi}_\textrm{{opt}}^\textrm{{GHZ}}=\alpha_{F}\ket{F,F}^{\otimes N}\!+\alpha_0\ket{F,0}^{\otimes N}\!+\alpha_{-F}\ket{F,-F}^{\otimes N},
\end{equation}
with $\alpha_F$, $\alpha_0$, and $\alpha_{-F}$ (satisfying $|\alpha_F|^2=|\alpha_{-F}|^2$ and $|\alpha_{0}|^2>0$) determined by hyperfine spin $F$.

Substituting the input state~\eqref{Eq:optimal-state-GHZ} into Eq.~\eqref{Eq:deltap_q_GHZ}, we get the corresponding measurement precision bounds for the two parameters
\begin{equation}\label{Eq:deltapoptimal}
 {\Delta p}_\textrm{{QCRB}}=\frac{1}{2\sqrt{2}N |\alpha_F| F},
\end{equation}
and
\begin{equation}\label{Eq:deltap2symme}
 {\Delta q}_\textrm{{QCRB}}=\frac{1}{2\sqrt{2}N |\alpha_F| F^2}.
\end{equation}
Similar with the optimal product state of individual atoms, ${\Delta p}_\textrm{{QCRB}}\propto\frac{1}{F}$ while ${\Delta q}_\textrm{{QCRB}}\propto\frac{1}{F^2}$.
However, the multimode GHZ state can improve the measurement precisions to the Heisenberg limit.
%
The log-log scaling of $\Delta p_\textrm{{QCRB}}$ and ${\Delta q}_\textrm{{QCRB}}$ versus the total atomic number $N$ for the optimal GHZ state and optimal product state are shown in Fig.~\ref{Fig_scaling_optimal_product_and GHZ}.
For GHZ state, the slopes are $-1$ which means the measurement precisions for the two parameters can both approach to the Heisenberg limit simultaneously, see Fig.~\ref{Fig_scaling_optimal_product_and GHZ}~(a) and (c).
While for product state, the slopes are just $-1/2$ and the measurement precisions for the two parameters can only reach the SQL simultaneously, see Fig.~\ref{Fig_scaling_optimal_product_and GHZ}~(b) and (d).
The comparison clearly shows that, the entanglement among the atoms in a multimode interferometer can improve the measurement precision of simultaneous multiparameter estimation.
%

\section{Simultaneous estimation versus individual estimation}\label{Sec4}

\begin{figure}[!htp]
  \includegraphics[width=1\columnwidth]{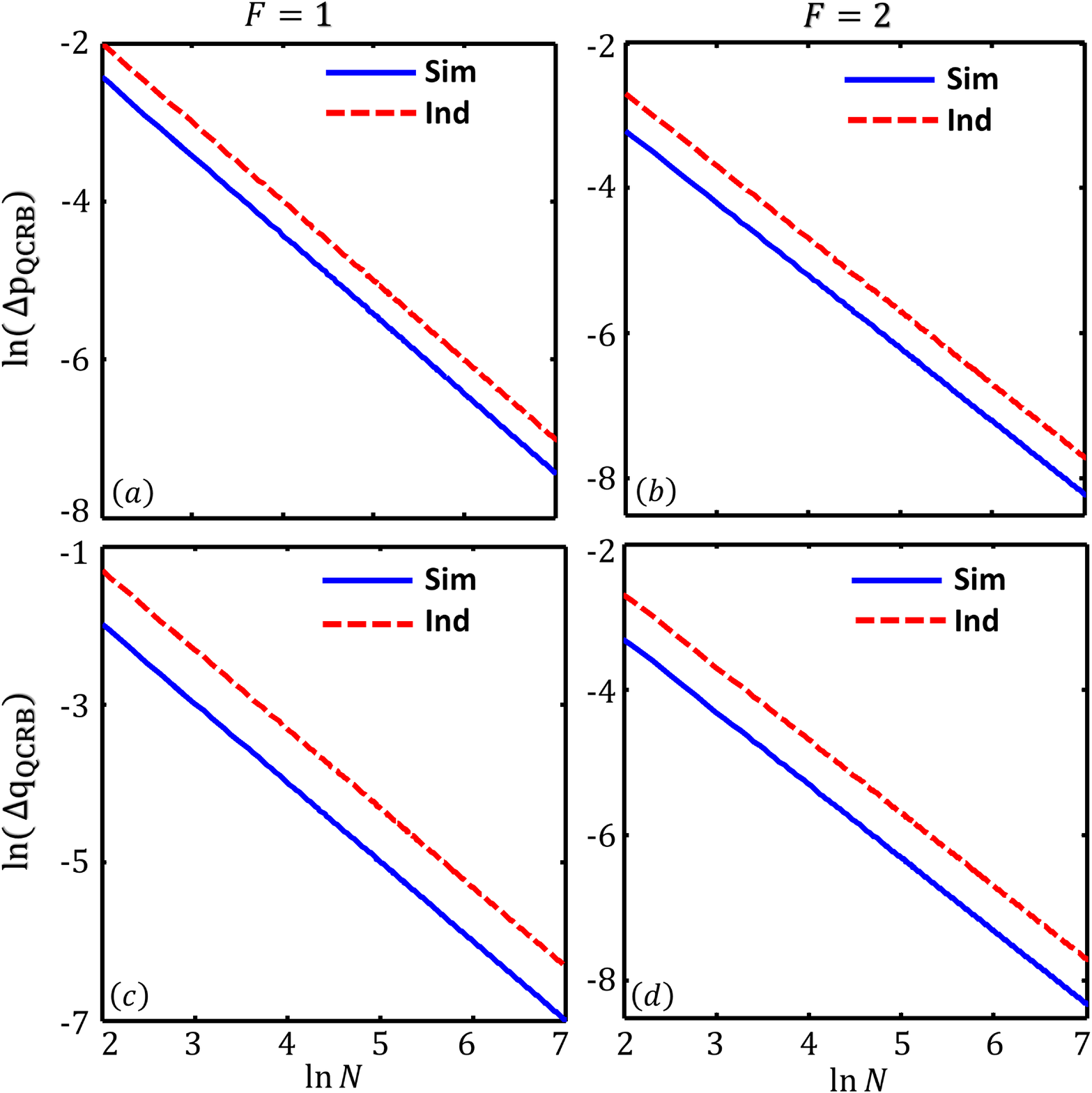}
  \caption{\label{Fig_Sim_vs_Indi}(color online).
  The log-log scaling of $\Delta {p_\textrm{{QCRB}}}$ and $\Delta {q_\textrm{{QCRB}}}$ versus the total atomic number $N$ for $F=1$ and $F=2$ under simultaneous estimation (blue solid lines) and individual estimation (red dashed lines).}
\end{figure}

In this section, we compare the measurement precisions via the simultaneous estimation and the individual estimation for the two parameters $p$ and $q$.
For the individual estimation, we evaluate one parameter independently with the other parameter assumed to be fixed.
The uncertainties of the two parameters within the individual estimation scheme can be given according to the conventional single parameter estimation theory.
That is,
\begin{eqnarray}\label{Eq:ind_deltap1_deltap2}
  \Delta^2p \ge {\Delta^2{p}}_\textrm{{QCRB}}^{\textrm{ind}}=\frac{1}{[\mathbf{F}_Q(p,q)]_{11}}, \nonumber\\
  \Delta^2q \ge {\Delta^2{q}}_\textrm{{QCRB}}^{\textrm{ind}}=\frac{1}{[\mathbf{F}_Q(p,q)]_{22}}. \nonumber\\
\end{eqnarray}
For the individual estimation, one can prepare different optimal input states to estimate $p$ and $q$ independently.
For the parameter $p$, the optimal GHZ state is in the form of $\ket{\Psi_p}_{\textrm{opt}}^\textrm{{GHZ}}=\frac{1}{\sqrt{2}}\ket{F,F}^{\otimes N}+\frac{1}{\sqrt{2}}\ket{F,-F}^{\otimes N}$, and the achievable measurement precision can be $\Delta p_{\textrm{QCRB}}^{\textrm{ind}}=\frac{1}{2NF}$.
On the other hand, for the parameter $q$, the optimal GHZ state becomes $\ket{\Psi_q}_{\textrm{opt}}^\textrm{{GHZ}}=\frac{1}{2}\ket{F,F}^{\otimes N}+\frac{1}{\sqrt{2}}\ket{F,0}^{\otimes N}+\frac{1}{2}\ket{F,-F}^{\otimes N}$, and the achievable measurement precision may be $\Delta q_{\textrm{QCRB}}^{\textrm{ind}}=\frac{1}{\sqrt{2}NF^2}$.
In comparison with the simultaneous estimation under the same resources, the total atomic number for the individual estimation should be $N'\rightarrow N/2$~\cite{PCHumphreys2013,TBaumgratz2016,SRagy2016} and the corresponding measurement precisions become $\Delta p_{\textrm{QCRB}}^{\textrm{ind}}=\frac{1}{NF}$ and $\Delta q_{\textrm{QCRB}}^{\textrm{ind}}=\frac{\sqrt{2}}{NF^2}$.

In Fig.\ref{Fig_Sim_vs_Indi}, we plot the log-log scaling of ${\Delta p}_\textrm{{QCRB}}$ and the ${\Delta p}_\textrm{{QCRB}}$ with respect to the total atomic number $N$ via the two different estimation schemes for $F=1$ and $F=2$.
It is shown that, the measurement precisions can both approach to the Heisenberg limit via simultaneous estimation as well as individual estimation.
However, the scaling lines for the simultaneous estimation are always lower than the ones for the individual estimation, which means that the simultaneous estimation scheme can yield better measurement precision than the individual estimation.

\section{Three-mode interferometry via Bose condensed spin-1 atoms\label{Sec5}}

In experiments, how to prepare the desired input states for sensing is another important problem.
Despite multimode GHZ states offer better interferometric advantages in theory, they are not easy to be realized in experiments.
How to prepare an available input state for implementing the simultaneous estimation becomes essential.
In this section, we discuss the three-mode interferometry with spin-1 Bose condensed atoms~\cite{YZou2018}.
For Bose condensed spin-1 atoms, spin exchange collision can be used to generate different kinds of input states.
We will illustrate how to prepare the three-mode entangled states via spin-mixing dynamics (SMD) and driving through quantum phase transition (QPT), and give their measurement precisions for simultaneously estimating the two parameters $p$ and $q$.
Then, we will also discuss how to estimate the two parameters via observable measurements.

In our below analysis, we consider the initial state with all $N$ atoms in $|F,m_F\rangle=|1,0\rangle$.
For an atom number conserved system, under spin exchange collision (by means of SMD or driving through QPT), its state will always be in the form of
\begin{equation}\label{Eq:Spin_one_magnetic_hamitonian}
\ket{\psi}_\textrm{{1}}=\sum\limits_{k=0}^{N/{2}}\alpha_k\ket{k,N-2k,k},
\end{equation}
with the Fock states $\ket{k,N-2k,k}$ denoting $k$ atoms in $|1,1\rangle$, $N-2k$ atoms in $|1,0\rangle$, and $k$ atoms in $|1,-1\rangle$.
Then, a $\frac{\pi}{2}$ pulse is applied for generating the input state,
i.e., $\ket{\psi}=\hat{U}_1\ket{\psi}_\textrm{{1}}$, $\hat{U}_1=\exp{[{\pi(\hat{a}^{\dagger}_{1}\hat{a}_{-1}-\hat{a}^{\dagger}_{-1}\hat{a}_{1})}/{4}]}$.
Implementing $\ket{\psi}$ for interrogation, two phase shifts $\varphi_1(p,q)$ and $\varphi_{-1}(p,q)$ relative to the mode of $\ket{F, m_F=0}$ are accumulated.
%
%
%
Thus the output state before measurement (we set the duration $T=1$) can be written as
\begin{widetext}
\begin{equation}\label{Eq:out_put_state}
\ket{\psi}_\textrm{{out}}=\sum\limits_{k=0}^{N/2}\alpha_k\sum\limits_{m=0}^{k}(-1)^{k-m}C_{k}^{m}e^{-i[(4m-2k)p+2kq]}\ket{2m,N-2k,2k-2m}.
\end{equation}
\end{widetext}
Here, the information of the parameters $p$ and $q$ are both imprinted in the output state.
The $\alpha_k $ and $C_{k}^{m}$ both are normalized coefficients~\cite{SDHuver2008}, the $C_{k}^{m}$ is given by
\begin{eqnarray}\label{Eq:Spin1-QCRB}
C_{k}^{m}\!\!=(-1)^{2k}\left[\left(
  \begin{array}{cc}
    2m \\
    m \end{array}\right)
  \left(
  \begin{array}{cc}
    2k-2m\\
    k-m \end{array}\right)\left({\frac{1}{2}}\right)^{2k}\right]^{\frac{1}{2}}.
\end{eqnarray}

Using Eqs.~\eqref{Eq:Cov-FQ}, ~\eqref{Eq:FQ} and~\eqref{Eq:Delta_p_q_general}, we obtain the simultaneous measurement precision bounds for $p$ and $q$, i.e.,
\begin{eqnarray}\label{Eq:general_evollution_QCRB}
  &&{\Delta p}_\textrm{{QCRB}}=\frac{1}{\sqrt{8\sum\limits_{k=0}^{N/2}{|\alpha_k|}^2({k+k^2})}},                                  \nonumber\\
  &&{\Delta q}_\textrm{{QCRB}}=\frac{1}{\sqrt{16\sum\limits_{k=0}^{N/2}|{\alpha_k}|^2 ({k^2})-{16\left({\sum\limits_{k=0}^{N/2}{|\alpha_k|}^2{k}}\right)}^2}}.\nonumber\\
\end{eqnarray}
This is the main result in this section. The measurement precisions ${\Delta p}_{QCRB}$ and ${\Delta q}_{QCRB}$ can be obtained according to the form of the prepared state $\ket{\psi_1}$.

\subsection{Three-mode interferometry via spin-mixing dynamics\label{Sec5a}}

The prepared state $\ket{\psi_1}$ can be generated by time evolution under spin mixing~\cite{TLHo1998,TOhmi1998,DMStamperKurn2013,MGabbrielli2015}.
The governed Hamiltonian is written as
\begin{equation}\label{H_SMD}
  \hat{H}_\textrm{SMD}=\kappa \left(\hat{a}^{\dagger}_{0}\hat{a}^{\dagger}_{0}\hat{a}_{1}\hat{a}_{-1}
  + \hat{a}_{0}\hat{a}_{0}\hat{a}^{\dagger}_{1}\hat{a}^{\dagger}_{-1}\right),
\end{equation}
where $\kappa$ is the strength of spin exchange collision and $\hat{a}_{m_F}$ denotes the annihilation operator for atoms in $\ket{1, m_F}$.
Initially with $\ket{\psi_0}=\ket{0,N,0}$, the prepared state can be obtained via SMD, i.e.,
\begin{equation}\label{psi_SMD}
  \ket{\psi_1}=e^{-i\hat{H}_\textrm{SMD}t}\ket{\psi_0}=e^{-i\hat{H}_\textrm{SMD}t}\ket{0,N,0}.
\end{equation}
Here, $t$ is the evolution time. For a fixed $\kappa$, different evolution time can result in different prepared state.
%
%
In our calculation, we set $\kappa=1$.
For a fixed total atom number $N$, one can obtained all the prepared states during the time evolution numerically.
Then, using Eq.~\eqref{Eq:general_evollution_QCRB}, we obtain the corresponding ${\Delta^2 p}_\textrm{{QCRB}}$ and ${\Delta^2 q}_\textrm{{QCRB}}$.
Minimizing ${\Delta^2 p}_\textrm{{QCRB}}+{\Delta^2 q}_\textrm{{QCRB}}$, we can obtain the optimal prepared state for simultaneously estimating $p$ and $q$.

\begin{figure}[!htp]
  \includegraphics[width=1\columnwidth]{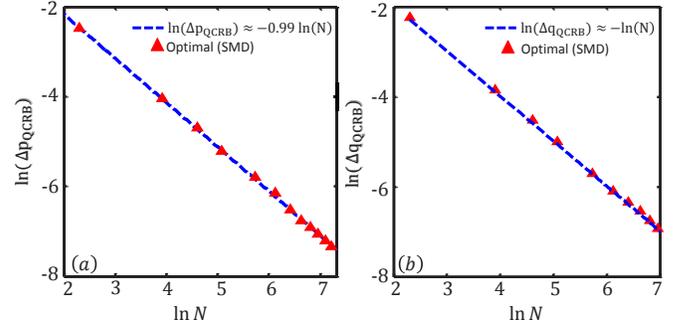}
  \caption{\label{Fig_Scaling_general_evolution}(color online).
  The log-log scaling of (a) $\Delta p_\textrm{{QCRB}}$ and (b) $\Delta q_\textrm{{QCRB}}$ versus the total atom number $N$ for the optimal prepared state via SMD (the red triangles). The blue dashed lines are the corresponding fitting results.}
\end{figure}

Although we cannot give the analytical form of the optimal prepared state for a given $N$, we numerically confirm that SMD could be an effective way for preparing the suitable states for simultaneous estimation.
We obtain the measurement precision $\Delta p_\textrm{{QCRB}}$ and $\Delta q_\textrm{{QCRB}}$ for the optimal prepared states under different total atom number $N$.
In Fig.~\ref{Fig_Scaling_general_evolution}, the log-log scaling of $\Delta p_\textrm{{QCRB}}$ and $\Delta q_\textrm{{QCRB}}$ with respect to $N$ is shown.
According to the fitting results, the slopes for $\Delta p_\textrm{{QCRB}}$ and $\Delta q_\textrm{{QCRB}}$ are both nearly $-1$.
Thus, by preparing the optimal state via SMD, the measurement precisions for estimating $p$ and $q$ can both approach the Heisenberg limit simultaneously.
Compared with the measurement precision bounds with multimode GHZ states, it preserves the Heisenberg-limited scalings but with a larger constant.
However, the state prepared via SMD may be much more experimentally feasible than the multimode GHZ states.

\subsection{Three-mode interferometry via driving through quantum phase transitions\label{Sec5b}}

\begin{figure}[!htp]
  \includegraphics[width=1\columnwidth]{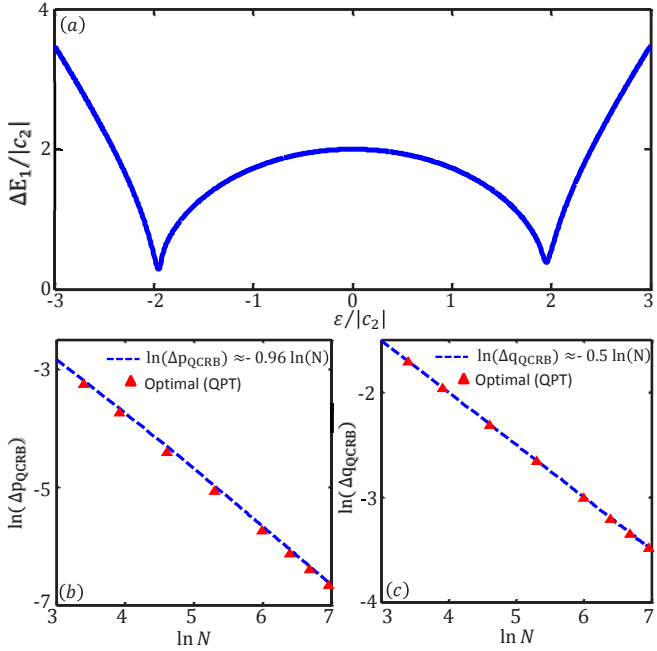}
  \caption{\label{Fig_Scaling_YouLi_optimal}(color online).
  (a) The gap $\Delta E_1$ in units of $|c_2|$ between the first excited and the ground state of Hamiltonian in Eq.~\eqref{Eq:Hamitonian_QPT} with respect to $\epsilon/|c_2|$, and $\epsilon/|c_2|=\pm2$ defines three distinct quantum phases.
  The log-log scaling of (b) $\Delta p_\textrm{{QCRB}}$ and (c) $\Delta q_\textrm{{QCRB}}$ versus the total atom number $N$ for the optimal prepared state via QPT (the red triangles). The blue dashed lines are the corresponding fitting results.}
\end{figure}

Driving through QPTs is another way for generating the desired input states.
Despite SMD can generate the prepared state, it sensitively depends on the control of evolution time and the states are invariably not steady.
In contrast, driving through QPTs can deterministically generate the prepared states.
Recently, entanglement generation by driving a $^{87}Rb$ spinor condensate QPTs has been realized in experiment~\cite{Luo2017}.
The evolution of the initial state $\ket{\psi_0}=\ket{0,N,0}$ is governed by the following Hamiltonian,
\begin{eqnarray}\label{Eq:Hamitonian_QPT}
  \hat{H}_{\textrm{QPT}}=\frac{c_2}{2N}[2(\hat{a}^{\dagger}_{0}\hat{a}^{\dagger}_{0}\hat{a}_{1}\hat{a}_{-1}
  +\hat{a}_{0}\hat{a}_{0}\hat{a}^{\dagger}_{1}\hat{a}^{\dagger}_{-1})\nonumber\\
         +(2\hat{N_0}-1)(N-\hat{N_0})]-\epsilon(t)\hat{N_0}.
\end{eqnarray}
Here, $|c_2|$ describes the rate of spin mixing process, $\epsilon=(\varepsilon_{+1}+\varepsilon_{-1})/2-\varepsilon_{0}$, with $\varepsilon_{m_F}$ is the energy of the $\ket{1, m_F}$,
and $\epsilon(t)$ can be tuned linearly with time in experiment.
The system possesses three distinct phases through the competition between $|c_2|$ and $q$~\cite{ZZhangi2013,Luo2017}.
For $\epsilon \gg 2|c_2|$, the ground state is polar state with all atoms in $\ket{1,0}$.
For $\epsilon \ll -2|c_2|$, the ground state becomes twin Fock state with atoms equally populated in $\ket{1,-1}$ and $\ket{1,1}$.
When $-2|c_2| < \epsilon < 2|c_2|$, the ground state corresponds to a superposition of all three components in the form of Eq.~\eqref{Eq:Spin_one_magnetic_hamitonian}.
If we ramp $\epsilon(t)$ from $\epsilon \gg 2|c_2|$ towards $\epsilon \ll -2|c_2|$ with very slow ramping rate (designed according to the energy spectrum in Fig.~\ref{Fig_Scaling_YouLi_optimal}~(a)), any instantaneous ground state can be adiabatically prepared.

Similarly, for a given total atom number $N$, we can obtained all the prepared states during the ramping numerically.
Using Eq.~\eqref{Eq:general_evollution_QCRB}, we obtain the corresponding ${\Delta^2 p}_\textrm{{QCRB}}$ and ${\Delta^2 q}_\textrm{{QCRB}}$.
Then, minimizing ${\Delta^2 p}_\textrm{{QCRB}}+{\Delta^2 q}_\textrm{{QCRB}}$, we can obtain the optimal prepared state for simultaneously estimating $p$ and $q$.
In Fig.~\ref{Fig_Scaling_YouLi_optimal}~(b) and (c), the log-log scaling of $\Delta p_\textrm{{QCRB}}$ and $\Delta q_\textrm{{QCRB}}$ with respect to $N$ is shown.
The optimal instantaneous prepared state are always in the region of $-2|c_2| < \epsilon < 2|c_2|$.
According to the fitting results, the slope for $\Delta p_\textrm{{QCRB}}$ is nearly $-1$ while the slope for $\Delta q_\textrm{{QCRB}}$ is only about $-0.5$.
That is, the measurement precision for $p$ can approach the Heisenberg limit but the measurement precision for $q$ can only attain the SQL.

%

\subsection{Simultaneous estimation via observable measurements\label{Sec5c}}

\begin{figure}[!htp]
  \includegraphics[width=1\columnwidth]{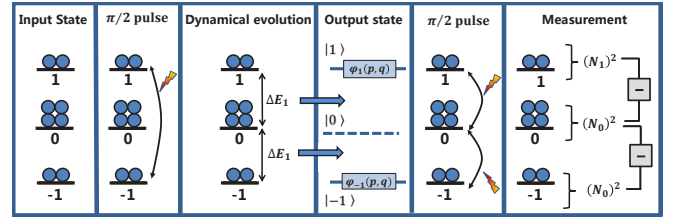}
  \caption{\label{Fig_Three_path_interferometry_scheme}(color online).
  Schematic of three-mode interferometry. Firstly, the system state among the three Zeeman sublevels is prepared. Secondly, a $\frac{\pi}{2}$ pulse coupling the states of $\ket{1,m_F=\pm1}$ is applied. Thirdly, interrogation for accumulating two relative phase $\varphi_1(p,q)$ and $\varphi_{-1}(p,q)$ is implemented. Fourthly, another two $\frac{\pi}{2}$ pulses coupling the three components are applied. Fifthly, the square of population difference measurement is used to extract the relative phase, and the two parameters $p$ and $q$ are derived from the relative phase.}
\end{figure}

\begin{figure}[!htp]
  \includegraphics[width=1\columnwidth]{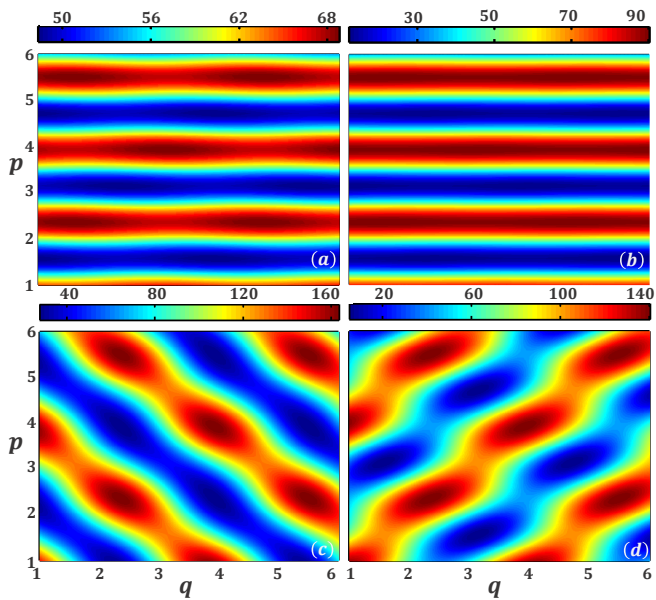}
  \caption{\label{Fig.9}(color online).
  Observable measurements under different values of $p$ and $q$.
  (a) and (c): $\langle{(\hat{N}_{1}-\hat{N}_{0})^2}\rangle$ versus $p$ and $q$.
  (b) and (d): $\langle{(\hat{N}_{-1}-\hat{N}_{0})^2}\rangle$ versus $p$ and $q$.
  (a) and (b): the prepared state $\ket{\psi_1}$ is the optimal one given in Sec.~\ref{Sec5a}.
  (c) and (d): the prepared state $\ket{\psi_1}$ is the optimal one given in Sec.~\ref{Sec5b}.
  Here, the total atomic number $N=20$.}
\end{figure}

Now, we turn to discuss how to extract the information of two parameters from observable measurements.
In practice, our three-mode interferometer can be implemented according to the procedures shown in Fig.~\ref{Fig_Three_path_interferometry_scheme}.
In our scheme, we evaluate the two parameters via measuring the square of population difference,
i.e., $\hat{\textbf{O}}=((\hat{N}_{1}-\hat{N}_{0})^2,(\hat{N}_{-1}-\hat{N}_{0})^2)$.
Before the square of population difference measurement, two $ \pi/2$ pulse is implemented to rotate the output state for recombination, i.e.,
$\ket{\psi}_{f}=\hat{U}_{22}\hat{U}_{21}\ket{\psi}_\textrm{{out}}$, where $\hat{U}_{21}=\exp{[{-i\pi(\hat{a}^{\dagger}_{1}\hat{a}_{0}+\hat{a}^{\dagger}_{0}\hat{a}_{1})}/{4}]}$
and $\hat{U}_{22}=\exp{[{-i\pi(\hat{a}^{\dagger}_{-1}\hat{a}_{0}+\hat{a}^{\dagger}_{0}\hat{a}_{-1})}/{4}]}$.
Applying the observable measurement $\hat{O}$ on the final state, one can obtain the expectation and standard deviation,
\begin{eqnarray}\label{Eq:Expection}
\langle\hat{O}\rangle_f=_f\!\langle \psi | {\hat{O}}|\psi \rangle_f,
\end{eqnarray}
\begin{eqnarray}\label{Eq:deviation}
(\Delta{\hat{O}})_f=\sqrt{ \langle {\hat{O}^2}\rangle_f-\langle {\hat{O}}\rangle_f^2}.
\end{eqnarray}
Following the above procedures, the expectations of the square of population difference measurement can be obtained,
\begin{eqnarray}\label{Eq:Nab^2}
&&\langle(\hat{N}_{1}-\hat{N}_{0})^2\rangle_f=C_{01}-\frac{1}{8}C_{1}\cos(4pt)\nonumber\\
&&+\textrm{R}\textrm{e}[C_{2}]\left[\frac{1}{4}\cos(-2pt+2qt)-\frac{9}{8}\cos(2pt+2qt)\right]\nonumber\\
&&+\textrm{I}\textrm{m}[C_{2}]\left[-\frac{1}{4}\sin(-2pt+2qt)-\frac{9}{8}\sin(2pt+2qt)\right]\nonumber\\
\end{eqnarray}
and
\begin{eqnarray}\label{Eq:Ncb^2}
&&\langle(\hat{N}_{-1}-\hat{N}_{0})^2\rangle_f=C_{02}-\frac{1}{2}C_{1}\cos(4pt)\nonumber\\
&&+\textrm{R}\textrm{e}[C_{2}]\left[\cos(-2pt+2qt)\nonumber-\textrm{I}\textrm{m}[C_{2}]
(-\frac{1}{4}\sin(-2pt+2qt)\right]\nonumber\\
\end{eqnarray}
The squares of the population  difference of the population differences are sensitive to the parameters $p$, $q$ and evolution time $t$,
where $C_{01}$, $C_{02}$, $C_{1}$, $C_{2}$ are the coefficients only dependent on the prepared state $\ket{\psi}_{1}$, i.e.,
\begin{eqnarray}\label{Eq:Cofficients}
C_{01}&=&\sum\limits_{k=0}^{N/2}|{\alpha_k}|^2({-\frac{9}{32}k}-\frac{57}{32}{k^2}+Nk+{\frac{11}{16}N}+{\frac{1}{16}N^2}),\nonumber\\
C_{02}&=&{\frac{1}{2}k}\sum\limits_{k=0}^{N/2}|{\alpha_k}|^2(N+2Nk-3k^2),\nonumber\\
C_{1}&=&\sum\limits_{k=0}^{N/2}|{\alpha_k}|^2(k+k^2),\nonumber\\
C_{2}&=&\sum\limits_{k=0}^{N/2-1}{\alpha^{*}_{k+1}}{\alpha_k}\sqrt{(N-2k-1)(N-2k)}(1+k).\nonumber\\
\end{eqnarray}
However, the form of the variance for the two observables both are complicated, we do not show them here.
The analytic results of $\langle(\hat{N}_{1}-\hat{N}_{0})^2\rangle$ and $\langle(\hat{N}_{-1}-\hat{N}_{0})^2)\rangle$ both display oscillatory patterns.
In our numerical simulation, we choose the optimal prepared states in Sec.~\ref{Sec5a} and Sec.~\ref{Sec5b}.
We show how $\langle(\hat{N}_{1}-\hat{N}_{0})^2\rangle$ and $\langle(\hat{N}_{-1}-\hat{N}_{0})^2)\rangle$ with respect to $p$ and $q$ for $N=20$, see Fig.~\ref{Fig.9}.

\begin{figure}[!htp]
  \includegraphics[width=1\columnwidth]{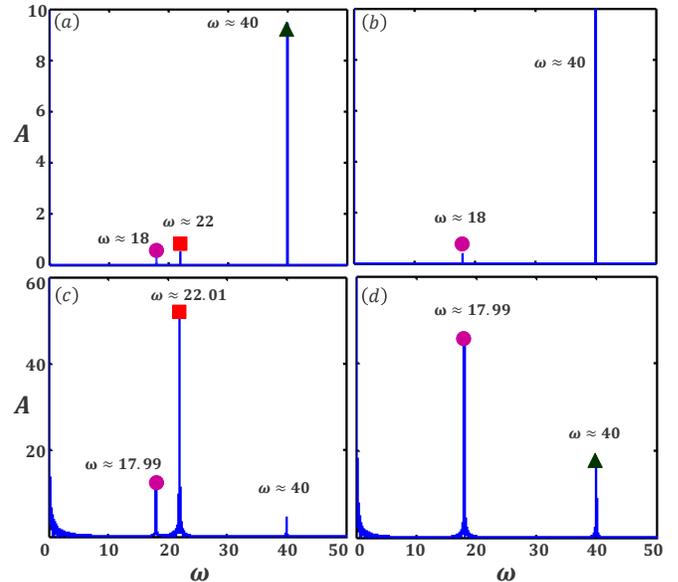}
  \caption{\label{Fig.10}(color online).
FFT spectrum of $\langle(\hat{N}_{1}-\hat{N}_{0})^2\rangle$ and $\langle(\hat{N}_{-1}-\hat{N}_{0})^2\rangle$.
  (a) and (c): The oscillation frequency of $\langle(\hat{N}_{1}-\hat{N}_{0})^2\rangle$ is close to $\omega=|4p|,|-2p+2q|,|2p+2q|$ (from right to left).
  (b) and (d): The oscillation frequency of $\langle(\hat{N}_{-1}-\hat{N}_{0})^2\rangle$ is close to $\omega=|4p|,|-2p+2q|$ (from right to left).
  (a) and (b): The prepared state $\ket{\psi_1}$ is the optimal one given in Sec.~\ref{Sec5a}.
  (c) and (d): The prepared state $\ket{\psi_1}$ is the optimal one given in Sec.~\ref{Sec5a}.
  Here, $N=20$, $p=10$ and $q=1$.}
\end{figure}

Then, we can obtain the values of $p$ and $q$ according to the fast Fourier transform (FFT).
The Fourier transform of $\langle{(\hat{N}_{1}-\hat{N}_{0})^2}\rangle$ clearly shows that oscillation frequencies are $\omega=|4p|,|2p+2q|,|-2p+2q|$.
While, the Fourier transform of $\langle{(\hat{N}_{-1}-\hat{N}_{0})^2}\rangle$ indicates that oscillation frequencies are $\omega=|4p|,|-2p+2q|$.
For practical situation, $|p| > |q|$, we have $|4p|>|2p+2q|>|-2p+2q|$.
%
%
%
In Fig.~\ref{Fig.10}, we show the FFT results for $p=10$, $q=1$.
It confirms that, the numerical results agree with theoretical predictions.
This implies that, one can estimate $p$ and $q$ simultaneously via observable measurements.
.

\section{conclutions\label{Sec6}}

In summary, we have presented a multimode quantum interferometry for multiparameter estimation with an ensemble of spinor atoms.
Based upon the proposed scheme, we have studied the quantum uncertainly bounds for simultaneous estimation of the linear zeeman coefficient and the quadratic zeeman coefficient.
We show that these quantum uncertainly bounds are dependent on both the hyperfine spin number $F$ and the total atomic number $N$.
Larger hyperfine spin number $F$ may offer better measurement precisions.
Especially, by employing the $N$-atom multimode GHZ state, the measurement precisions of the two parameters can simultaneously attain the Heisenberg limit, and they depend on both the hyperfine spin number $F$ and the total atomic number $N$ according to $\Delta p \propto 1/(FN)$ and $\Delta q \propto 1/(F^2N)$.
For a given total atomic number, we found that the measurement precisions from simultaneous estimation are better than the ones from individual estimation.

Through taking a three-mode interferometry with Bose condensed spin-1 atoms for example, we give the achievable measurement precision bounds for the desired input states via SMD and driving through QPTs.
We found that the measurement precisions for estimating $p$ and $q$ may simultaneously approach the Heisenberg limit.
Furthermore, we show how to simultaneously estimate $p$ and $q$ via observable measurements.
Our proposed scheme may point out a paradigm for implementing multiparameter estimation via multimode quantum correlated states.

\section*{Acknowledgements}
This work is supported by the National Natural Science Foundation of China (NNSFC) under Grants No. 11574405 and 11704420. J. H. is partially supported by National Postdoctoral Program for Innovative Talents of China (BX201600198).

\end{document}